\documentclass[aps,pre,12pt,groupedaddress,preprintnumbers,
superscriptaddress,nofootinbib]{revtex4-1}

\usepackage{amsmath,amssymb}
\usepackage[breaklinks=true,colorlinks=true,linkcolor=blue,citecolor=blue]{hyperref}
\usepackage[dvipdfmx]{graphicx}
\usepackage{xcolor}
\usepackage{comment}
\usepackage{setspace}

\newcommand{\vx}{\vec{x}}
\newcommand{\vy}{\vec{y}}
\newcommand{\cA}{{\cal A}}
\newcommand{\hA}{{\hat A}}
\newcommand{\hB}{{\hat B}}
\newcommand{\ha}{{\hat a}}
\newcommand{\hb}{{\hat b}}
\newcommand{\hc}{{\hat c}}
\newcommand{\hd}{{\hat d}}
\newcommand{\tF}{{\tilde F}}

\newcommand{\be}{\begin{eqnarray}}
\newcommand{\ee}{\end{eqnarray}}
\newcommand{\tr}{\mathrm{tr\,}}
\newcommand{\nt}{\notag\\}

\begin{document}

\preprint{KEK-TH-2205}

\title{Current Algebra Formulation of Quantum Gravity\\ and Its Application to Cosmology}

\author{Shotaro Shiba Funai}
\email{shotaro.funai@oist.jp}
\affiliation{Physics and Biology Unit, Okinawa Institute of Science and Technology (OIST), 
  1919-1 Tancha Onna-son, Kunigami-gun, Okinawa 904-0495, Japan}

\author{Hirotaka Sugawara}
\email{sugawara@post.kek.jp}
\affiliation{High Energy Accelerator Research Organization (KEK),
   1-1 Oho, Tsukuba, Ibaraki 305-0801, Japan}

\begin{abstract}
\vspace*{10pt}
Gravity theory based on current algebra is formulated. 
The gauge principle rather than the general covariance combined with 
the equivalence principle plays the pivotal role in the formalism{, and}
the latter principles are derived as a consequence of the theory. 
In this approach,
it turns out that gauging the Poincar\'e algebra is not appropriate 
but gauging the $SO(N,M)$ algebra gives a consistent theory.
{This makes it} 
possible to have Anti-de Sitter 
and de Sitter 
space-time by adopting a relation 
between the spin connection and the tetrad field. 
The Einstein equation is a part of our basic equation for gravity 
which is written {in terms of} 
the spin connection. 
When this formalism is applied to the $E(11)$ {algebra} in which the three-form 
antisymmetric tensor is a part of gravity multiplet, 
we have a 
current algebra gravity theory based on M-theory to be applied to cosmology
in its classical limit. 
Without introducing any other ad-hoc field, we can obtain accelerating universe 
in the manner of 
the ``inflating'' universe at its early stage. 
\end{abstract}

\maketitle

\section{Introduction}

There is a long history of formulating gravity as a gauge theory starting from 
a pioneering work of Uchiyama~\cite{ref1}.\footnote{
See also Ref.\,\cite{ref1a}. They are followed by Refs.\,\cite{ref1b,ref1c,ref1d,ref1e,ref1f,ref1g}.}
In many cases, such attempts start with the Einstein-Hilbert action and 
rewrite it in terms of spin connection and tetrad both of which are vector fields 
rather than tensor, thus making the theory a vector gauge theory. 
This approach is not satisfactory 
{in that}, 
though the resulting theory is a vector gauge theory,
the general covariance and the equivalence principle 
on which the Einstein theory is based
are more fundamental than 
the gauge principle on which the standard model and its extensions are based. 
Another  question remains  whether the theory can be quantized,
since the original action is the same as the Einstein-Hilbert. 

Here we present a completely different approach to this problem, 
which we call the current algebra formulation of the gravity. 
We do not assume the Einstein-Hilbert action but  eventually derive 
the Einstein equation in its quantum form based on the gauge principle 
but neither on general covariance nor on equivalence principle. 
Our formulation is quantum from the beginning and all the fields are $q$-numbers,
although we apply our formalism to the classical case of cosmology. 

In another aspect, M-theory was formulated by P. West 
using the Kac-Moody algebra $E(11)$~\cite{ref2,ref2a,ref2b,ref2c,ref2d}.
We followed his idea and combined it with the current algebra formulation~\cite{ref3,ref3a}.
Although the current algebra formulation is general 
and does not have to be combined with $E(11)$ formulation of the M-theory, 
it turns out that we obtain interesting cosmological results in this case. 

The next question is: why the current algebra formulation 
rather than the usual quantum theory formalism with $E(11)$ symmetry?  
To answer this question,  we refer to the work done in 1968 by Bardacki, Frishman and Halpern~\cite{ref4}.

First remember that, to get the string theory from $SU(N)$ {gauge theory}, 
we need {to take the limit} $N\to\infty$ with $g^2 N$ fixed.
Bardacki, Frishman and Halpern~\cite{ref4}
proved that the massive Yang-Mills theory becomes a current algebra theory 
with current-current energy momentum tensor, if we take the limit
$g\to 0$ and $m\to 0$.
This implies that the large $N$ limit of massive $SU(N)$ gauge theory becomes 
the current-current theory, if we take
$m=\frac{m_0}{N}$ and $N\to\infty$ with $m_0$ and $g^2 N$ fixed.
Therefore, the $N\to\infty$ limit of $SU(N)$ gauge theory is in fact 
the current-current theory presumably with some kind of Kac-Moody symmetry.
This concludes the {reason we} 
use the $E(11)$ Kac-Moody current-current theory 
to describe the M-theory.

The present work is in a way to clarify and correct 
some of the concepts used in {our previous studies}~\cite{ref3,ref3a} and
{to} apply the theory formulated in this scheme to cosmology. 
We first review the current algebra formulation of gravity theory 
not necessarily restricted to $E(11)$ algebra but  based on $SO(N,M)$ algebra,
and make some changes in the assumption 
in the original formulation~\cite{ref3,ref3a}.

We start with the current algebra formulation based on general (Lie or Kac-Moody) algebra.
To each generator $G^A$ of the algebra $\cal A$,
there {is the corresponding} current $J^A_\mu(x)$ 
that satisfies the commutation relations~\cite{ref5}:
\begin{eqnarray}
\left[ J_0^A(x), J_0^B(y) \right] \big|_{x_0=y_0}
 &=& i f^{ABC} J_0^C \delta(\vx - \vy) \label{eq1}\\
\left[ J_0^A(x), J_n^B(y) \right] \big|_{x_0=y_0}
 &=& i f^{ABC} J_n^C \delta(\vx - \vy)
       + iC\delta^{AB}\partial_n \delta(\vx-\vy)  \label{eq2}\\
\left[ J_m^A(x), J_n^B(y) \right] \big|_{x_0=y_0}
 &=& 0\,. \label{eq3}
\end{eqnarray}
for $m,n \neq 0$. 
Here, $x, y$ denote the space-time components of local Lorentzian frame 
of $d$-dimensional Riemannian space-time. 
The indices $\mu,\nu,\ldots$ run from $0$ to $d-1$, 
and $m,n,\ldots$ run from $1$ to $d-1$. 
$\vx$ and $\vy$ designate {the spatial} 
components of $x$ and $y$, respectively. 
$f^{ABC}$ is the structure constant of the algebra $\cA$ and is given by
\be
\left[ G^A, G^B \right] = i f^{ABC} G^C, \qquad
\tr(G^A G^B) = \delta^{AB}.
\ee
To understand the current algebra,
{we can regard it} 
as an extension of Heisenberg commutation relation 
(by the term $iC\delta^{AB}\partial_n(\vx-\vy)$ in Eq.\,(\ref{eq2})) 
to include the gauge symmetry algebra or vice versa.
The equation of motion for the current $J^A_\mu(x)$ is given 
by the energy momentum tensor:
\be
\Theta_{\mu\nu}(x)
 &=& \frac{1}{C} \left[ \tr(\Omega_\mu \Omega_\nu) 
        - \frac12 \eta_{\mu\nu} \tr(\Omega_\rho \Omega^\rho) \right] \notag\\
 &=& \frac{1}{C} \left[ J^A_\mu(x) J^A_\nu(x) 
        - \frac12 \eta_{\mu\nu} \eta^{\rho\sigma} J^A_\rho(x) J^A_\sigma(x) \right]
\ee
where we take the local Lorentz frame
$g_{\mu\nu}\to\eta_{\mu\nu} = {\rm diag.} (-1,1,\ldots,1)$ and 
\be
\Omega_\mu(x) = J^A_\mu(x) G^A.
\ee
Note that, in our notation, duplicate indices $A,B,\ldots$ are summed even if both of them are upper (or lower) indices.

The energy momentum tensor is constructed to satisfy 
the Schwinger commutation relation~\cite{ref6} 
which is the manifestation of local Lorentz invariance:
\be \label{eq7}
\left[\Theta_{00}(x), \Theta_{00}(y)\right]\big|_{x_0=y_0}
 = -i \left( \Theta_{0m}(x) + \Theta_{0m}(y) \right) \partial_m \delta(\vx-\vy)\,.
\ee
The equation of motion is {given as}
\be \label{eq8}
-i\partial_\mu J^A_\nu(x) = \left[ P_\mu, J^A_\nu(x) \right]
\ee
where
\be \label{eq9}
P_\mu := \int \Theta_{0\mu}(x) d\vx\,.
\ee
Then we obtain the following two equations from Eqs.\,(\ref{eq8}) and (\ref{eq9}):
\be \label{eq10}
F^A_{\mu\nu} := D^{AB}_\mu J^B_\nu(x) - D^{AB}_\nu J^B_\mu(x) = 0
\ee
with 
\be \label{eq11}
D^{AB}_\mu := \delta^{AB}\partial_\mu - \frac{i}{2C}f^{ABC} J_\mu^C
\ee
and
\be \label{eq12}
\partial_\mu J^{\mu A} = 0\,.
\ee
The most important interpretation of Eqs.\,(\ref{eq10})--(\ref{eq12}) 
for our purpose is that 
we {can} regard the current $J^A_\mu(x)$ as the gauge field itself. 
Then, we observe that Eq.\,(\ref{eq10}) is gauge covariant with the usual gauge transformation based on the algebra $\cA$. 
With this observation, Eq.\,(\ref{eq12}) is the gauge fixing equation (Landau gauge) rather than the dynamical equation. 
For this interpretation to be valid, it is important to realize that the affine connection
$\Gamma^\nu_{\mu\lambda}$ in the local Lorentz frame satisfies
\be
\eta^{\mu\lambda} \Gamma^\nu_{\mu\lambda} = 0\,.
\ee
Therefore, we obtain
\be
&& D_\mu A^\mu = D_\mu g^{\mu\rho} A_\rho = g^{\mu\rho} D_\mu A_\rho \notag\\
&& \xrightarrow{\text{Local Lorentz frame}}\quad
\eta^{\mu\rho} D_\mu A_\rho
= \left(\eta^{\mu\nu}\partial_\mu - \eta^{\mu\rho} \Gamma^\nu_{\mu\rho} \right)
   A_\nu
= \partial_\mu A^\mu\,.
\ee
Under this interpretation, we can ignore Eq.\,(\ref{eq12}) 
if we want to use some other gauge fixing.  
The only  dynamical equation we have is, therefore, Eq.\,(\ref{eq10}). 

This interpretation becomes important 
when we apply the above method with $\cA$ to be $SO(N,M)$ or $E(11)$. 
In these cases, Eq.\,(\ref{eq10}) is gauge covariant and also general covariant 
if we assume the affine connection $\Gamma^\nu_{\mu\lambda}$
to be symmetric in $\mu$ and $\lambda$, i.e., the torsion-less condition.

In fact, all these geometric variables such as the metric tensor $g_{\mu\nu}$, 
the affine connection $\Gamma^\nu_{\mu\lambda}$,
and the curvature tensor $R^\mu{}_{\nu\kappa\lambda}$
will be defined later in terms of the gauge fields which appear in Eq.\,(\ref{eq10}).  

Since in writing Eq.\,(\ref{eq7}) we take the local Lorentz frame,
it is intriguing that we can obtain Eq.\,(\ref{eq10})
that is {\em automatically} general covariant. 
In this sense the general covariance is not an input of the theory 
but the consequence of the gauge principle formulated in the current algebra. 
We will show later that the equivalence principle can be also satisfied 
under a certain condition, {which means that} 
we expect some possibility of its violation when this condition is not satisfied.

{In this paper,} we apply the above formalism to
\begin{enumerate}
\item
$\cA=SO(N,1)$ corresponding to pure gravity with positive cosmological constant, and 
$\cA=SO(N,2)$ which describes the pure gravity with negative cosmological constant
\item
$\cA=E(11)$ with negative and positive cosmological constant {in} 
the compactified space and our $(3,1)$ space-time, respectively.
{(As we will see in Sec.\,\ref{sec:5}, the cosmological constant is in fact a time-dependent function.)}
\end{enumerate}
We also discuss the cosmological implication in the latter case.

\section{Current algebra for $\cA=SO(N,1)$ or $SO(N-1,2)$}
\label{sec:2}

The generators of the $SO(N,1)$ algebra are 
$M_{ab}$ (where $a,b=0,1,\ldots,N$)
and their commutation relations are 
\be \label{comm1}
{}[M_{ab},M_{cd}] &=& 
 \eta_{ad}M_{bc} - \eta_{ac}M_{bd} -\eta_{bd}M_{ac} + \eta_{bc}M_{ad} \,.
\ee
The currents corresponding to these generators are defined as
\be \label{curr1}
\Omega_\mu(x) = J_\mu^A(x) G^A = 
K_\mu^{ab}(x) M_{ab}\,.
\ee
Then, using Eqs.\,(\ref{eq1})--(\ref{eq3}), 
the current algebra based on $\cA=SO(N,1)$ is given by\footnote{
Note that $J_\mu^A$ here includes an antisymmetric tensor $K_\mu^{ab}$.
The structure constant $f^{(ab)(cd)}{}_{(ef)} = \eta^{ad}\delta^b_{[e}\delta^c_{f]} + \cdots$
is read off from the commutation relation $[M_{ab},M_{cd}]$.
The $\delta^{(ab)(cd)}$ is defined as $\eta^{ac}\eta^{bd} - \eta^{ad}\eta^{bc}$,
since the indices $a,b,\ldots$ denote directions in the $(N,1)$ flat space-time.
When we discuss another algebra $\cA$ where $J_\mu^A$ includes higher-rank
totally antisymmetric tensors, we can define them in a similar way.}
\be \label{eq15}
\left[K_0^{ab}(x),K_0^{cd}(y)\right]\big|_{x_0=y_0} &=& 
\left(\eta^{ad}K_0^{bc}-\eta^{ac}K_0^{bd}-\eta^{bd}K_0^{ac}+\eta^{bc}K_0^{ad}\right)\delta(\vx-\vy) \nt
\left[K_0^{ab}(x),K_m^{cd}(y)\right]\big|_{x_0=y_0} &=& 
\left(\eta^{ad}K_m^{bc}-\eta^{ac}K_m^{bd}-\eta^{bd}K_m^{ac}+\eta^{bc}K_m^{ad}\right)\delta(\vx-\vy) \nt
&& +\, iC \left(\eta^{ac}\eta^{bd}-\eta^{ad}\eta^{bc}\right)\partial_m\delta(\vx-\vy)\nt
\left[K_m^{ab}(x),K_n^{cd}(y)\right]\big|_{x_0=y_0} &=& 0\,.
\ee
The resulting equations of motion are
\be\label{eq18}
D^{cd,ab}_\mu K_{\nu ab}(x) - D^{cd,ab}_\nu K_{\mu ab}(x) = 0\,,\quad
\partial^\mu K_\mu^{ab}(x) = 0
\ee
where the covariant derivative is {defined as}
\be\label{eq19}
D^{cd,ab}_\mu := \eta^{ca}\eta^{db}\partial_\mu +\frac{i}{2C}\left(
 \eta^{bc}K_\mu^{da}-\eta^{da}K_\mu^{bc}\right).
\ee
By defining the spin connection
\be
\omega_\mu^{ab}(x) := \frac{{i}}C K_\mu^{ab}(x)\,,
\ee
Equation (\ref{eq18}) becomes
\be\label{eq21}
\partial_\mu \omega_\nu^{cd}(x) - \partial_\nu \omega_\mu^{cd}(x) 
+\omega_{\mu a}^c(x) \omega_\nu^{ad}(x) 
-\omega_{\nu a}^c(x) \omega_\mu^{ad}(x) 
=0\,.
\ee
The next question is how we obtain the tetrad $e_\mu^a$ which is necessary to provide a geometrical interpretation to 
Eq.\,(\ref{eq21}). 
We follow Stelle and West~\cite{ref7} for this purpose and define
\be\label{eq22}
\omega_\mu^{\ha N}(x) =: \kappa e_\mu^\ha(x)
\ee
where $\ha=0,1,\ldots,N-1$.
This means that Eq.\,(\ref{curr1}) is rewritten as
\be
\Omega_\mu(x) = J_\mu^A(x) G^A 
= e_\mu^\ha(x) P_\ha + K_\mu^{\ha\hb}(x) M_{\ha\hb}
\ee
where the translation is defined as $P_\ha := -i\kappa C M_{\ha N}$.
The difference  from the  case of Stelle-West~\cite{ref7} is that we are treating $\cA = SO(N, 1)$ and the Stelle-West is concerned about $SO(3, 2)$. 
Consequently, 
we obtain the positive cosmological constant (de Sitter space), 
while Stelle-West obtains the negative one (Anti-de Sitter space). 
Then Eq.\,(\ref{eq21}) becomes
\be\label{eq23}
&&
\partial_\mu \omega_\nu^{\hc\hd}(x) - \partial_\nu \omega_\mu^{\hc\hd}(x) 
+\omega_{\mu \ha}^\hc(x) \omega_\nu^{\ha\hd}(x) 
-\omega_{\nu \ha}^\hc(x) \omega_\mu^{\ha\hd}(x)  
\nt &&
=\kappa^2 \left(
 \omega_{\nu N}^\hc(x) \omega_\mu^{N\hd}(x)
- \omega_{\mu N}^\hc(x) \omega_\nu^{N\hd}(x) 
\right)
=\kappa^2\left(e_\mu^\hc(x)e_\nu^\hd(x) - e_\nu^\hc(x) e_\mu^\hd(x)\right)
\ee
and we also have
\be\label{eq24}
\partial_\mu e_\nu^\hc(x) - \partial_\nu e_\mu^\hc(x) 
+ \omega^\hc_{\mu \ha}(x) e_\nu^\ha(x)
- \omega^\hc_{\nu \ha}(x) e_\mu^\ha(x)
= 0\,.
\ee
This can be simply rewritten as
\be\label{eq25}
D^\hc_{\mu \ha} e_\nu^\ha(x) - D^\hc_{\nu \ha} e_\mu^\ha(x) = 0
\ee
where 
$D^\hc_{\mu \ha} := \delta_\ha^\hc \partial_\mu + \omega^\hc_{\mu \ha}$.

If we adopt $\cA = SO(N-1, 2)$ and define the tetrad 
just as in Eq.\,(\ref{eq22}):
$\omega_\mu^{\ha N} = \kappa e_\mu^\ha(x)$,
where 
$N$ stands for one of the directions corresponding to the negative metric sign, 
all we obtain is the same as above except 
an overall negative sign in 
the right hand side of Eq.\,(\ref{eq23}).
{Thus we have} 
the Anti-de Sitter space,
{just as in the Stelle-West's case}.

The significance of the Stelle-West ansatz~\cite{ref4} is that it gives the origin of the cosmological constant as the coefficient of the equation which relates the spin connection to the tetrad. 
We will generalize this {coefficient} later as the time-dependent dynamical 
{field} 
and apply it to cosmology. 

\subsection{Geometric interpretation}

To give the  geometric interpretation to our formalism, we define 
1. space-time metric, 2. affine connection,  and 3. curvature. 
We follow the conventional definition in each case.

\begin{enumerate}
\item The metric is defined as
\be
g_{\mu\nu} := e_\mu^a e_{\nu a}
\ee
where the tetrad $e_\mu^a$ is defined in terms of the spin connection as in Eq.\,(\ref{eq22}).
{Note that the indices $a,b,\ldots = 0,1,\ldots,N-1$ here.}

\item The affine connection is defined as
\be
\Gamma_{\mu\lambda}^\nu := e_a^\nu \left(\partial_\mu e_\lambda^a + 
 \omega^a_{\mu b} e_\lambda^b \right)
\ee
with $e_a^\nu e^a_\mu = \delta_\mu^\nu$.
As is well known, this is equivalent to
\be
\tilde D_{\mu a}^c e_\nu^a 
:= D_{\mu a}^c e_\nu^a - \Gamma_{\mu\lambda}^\sigma e_\sigma^c = 0\,.
\ee
Equation 
(\ref{eq25}) is trivially satisfied by this definition of the affine connection, 
when we have the torsion-less condition:
\be 
\Gamma_{\mu\lambda}^\sigma = \Gamma_{\lambda\mu}^\sigma\,.
\ee

\item The curvature is defined to be
\be\label{eq31}
R^\lambda{}_{\sigma\mu\nu}
&:=& e_c^\lambda e_\sigma^d R^c{}_{d\mu\nu}
= e_c^\lambda e_\sigma^d \left(
 \partial_\mu\omega_{\nu d}^c - \partial_\nu\omega_{\mu d}^c
 -\omega_{\mu a}^c\omega_{\nu d}^a + \omega_{\nu a}^c\omega_{\mu d}^a
 \right) \nt
&=& e_c^\lambda e_{\sigma d}\left(
 D^{cd,ab}_\mu \omega_{\nu ab} - D^{cd,ab}_\nu \omega_{\mu ab}
 \right).
\ee
{Although $D_\mu^{cd,ab}$ is defined in Eq.\,(\ref{eq19}),}
we sum over only $a,b,\ldots=0,1,\ldots,N-1$ here,
since the direction $N$ is used to define the tetrad. 
As is well known, we can write the curvature $R^\lambda{}_{\sigma\mu\nu}$ entirely  
in terms of the affine connection $\Gamma_{\mu\lambda}^\sigma$ or the metric $g_{\mu\nu}$ using above definitions. 
Using Eqs.\,(\ref{eq22}) and (\ref{eq31}), we obtain
\be
R^\lambda{}_{\sigma\mu\nu}
=\pm \kappa^2 e_c^\lambda e_{\sigma d}\left(
 e_\mu^c e_\nu^d - e_\nu^c e_\mu^d \right)
\ee
where $+$ for $SO(N,1)$ and $-$ for $SO(N-1,2)$.
The Einstein equation is just the trace of this equation over $\lambda$ and $\mu$, then
\be
R_{\sigma\nu} := R^\mu{}_{\sigma\mu\nu}
=\pm \kappa^2 e_c^\mu e_{\sigma d}\left(
 e_\mu^c e_\nu^d - e_\nu^c e_\mu^d \right)
=\pm \kappa^2 (N-1)  g_{\sigma\nu}\,.
\ee
\end{enumerate}

\section{Current algebra for $\cA=E(11)$}
\label{sec:E11}

The case of $\cA=E(11)$ corresponds to the M-theory as initiated by P.~West~\cite{ref2,ref2a,ref2b,ref2c,ref2d}. 
The essential difference from the above 
cases {of $SO(N,1)$ and $SO(N-1,2)$} is that the regular representation of $E(11)$ contains not just the spin connection but the matter field/current,
notably the $B_\mu^{abc}$ field/current
corresponding to the three-form 
antisymmetric tensor field $C_{\mu\nu\rho}$ 
that appears in the low energy action of M-theory. 
$B_\mu^{abc}$ is a part of gravity multiplet belonging to the $E(11)$ regular representation just like spin connection, but it is a matter field from the viewpoint of $SO(10,1)$ 
{subgroup of $E(11)$, i.e., the Lorentz group in the $(10,1)$ space-time 
of M-theory.}
The precise relation between $B_\mu^{abc}$ and $C_{\mu\nu\rho}$ 
{is given in Ref.\,\cite{ref3a}.}

Moreover, in addition to the $E(11)$ regular representation, 
we need another representation which is called ``vector" representation $\ell(1)$. 
We assign the tetrad field/current to this representation. 
This {seems} 
in a way inconsistent with the idea of regarding the tetrad as a part of the spin connection described 
{in the previous section, and we will discuss it later.}

Therefore, we use two kinds of the representation of $E(11)$ 
to define the currents. 
One is the infinite dimensional regular representation which includes 
the $SO(10,1)$ generators $M_{ab}$ and the antisymmetric tensor operator {$S_{abc}$}.
The commutation relations among them are 
\be
{}[M_{ab},M_{cd}] &=& 
 \eta_{ad}M_{bc} - \eta_{ac}M_{bd} -\eta_{bd}M_{ac} + \eta_{bc}M_{ad} \nt
{}[M_{ab},S_{cde}] &=& 
 3\left(\eta_{b[c}S_{de]a} - \eta_{a[c}S_{de]b}\right) \nt
{}[S^{abc},S_{def}] &=&
 -36\delta^{[a}_{[d}\delta^b_e M^{c]}_{f]}\,.
\ee
The corresponding currents are defined as
\be \label{curr2}
\Omega_\mu(x) = J_\mu^A(x) G^A
= 
K_\mu^{ab}(x) M_{ab} + B_\mu^{abc}(x) S_{abc}
\ee
and  
the corresponding fields/currents 
satisfy the following commutation relations~\cite{ref3,ref3a}:
\be
\left[K_0^{ab}(x),K_0^{cd}(y)\right]\big|_{x_0=y_0} &=& 
2 \left(\eta^{a[c}K_0^{d]b}-\eta^{b[c}K_0^{d]a}\right)\delta(\vx-\vy) \nt
\left[K_0^{ab}(x),K_m^{cd}(y)\right]\big|_{x_0=y_0} &=& 
2 \left(\eta^{a[c}K_m^{d]b}-\eta^{b[c}K_m^{d]a}\right)\delta(\vx-\vy) 
+ 2iC \eta^{a[c}\eta^{d]b} \partial_m\delta(\vx-\vy)\nt
\left[K_m^{ab}(x),K_n^{cd}(y)\right]\big|_{x_0=y_0} &=& 0
\ee
\be
\left[B_0^{abc}(x),B_{0,def}(y)\right]\big|_{x_0=y_0} &=& 
{-36}\delta^{[a}_{[d}\delta^b_eK_{0,f]}^{c]}\delta(\vx-\vy) \nt
\left[B_0^{abc}(x),B_{m,def}(y)\right]\big|_{x_0=y_0} &=& 
-36\delta^{[a}_{[d}\delta^b_eK_{m,f]}^{c]}\delta(\vx-\vy)
+ 6iC\delta^{a}_{[d}\delta^b_e\delta^{c}_{f]}\partial_m \delta(\vx-\vy)\nt
\left[B_m^{abc}(x),B_{n,def}(y)\right]\big|_{x_0=y_0} &=& 0
\ee
\be
\left[B_0^{abc}(x),K_\mu^{de}(y)\right]\big|_{x_0=y_0} &=&
3\left(\eta^{d[a} B_\mu^{bc]e} - \eta^{e[a} B_\mu^{bc]d}\right)\delta(\vx-\vy)\nt
\left[K_0^{de}(x),B_\mu^{abc}(y)\right]\big|_{x_0=y_0} &=& 
-3\left(\eta^{d[a} B_\mu^{bc]e} - \eta^{e[a} B_\mu^{bc]d}\right)\delta(\vx-\vy)\nt
\left[B_m^{abc}(x),K_n^{de}(y)\right]\big|_{x_0=y_0} &=& 0\,. 
\ee
The notation is such that 
$\mu,\nu,\ldots = 0, 1, \ldots, 10$ and 
$m, n, \ldots = 1, \ldots, 10$ {for the curved space-time},
while $a, b, \ldots$ are the flat space indices which run over $0, 1, \ldots, 10$. 
We also note that the constant $C$ has the dimension of (length)${}^{-9}$ and the fields/currents $K_\mu^{ab},B_\mu^{abc},e_\mu^a,\ldots$ have dimension (length)${}^{-10}$.

In addition, we define the currents which correspond to the operators belonging to 
the $\ell(1)$, or the vector representation of $E(11)$. They include the translation $P_a$
and the commutation relations are 
\be
{}[M_{ab}, P_c] &=&
 \eta_{cb}P_a - \eta_{ca}P_b \nt 
{}[S_{abc},P_d] &=& 0\,.
\ee
Then the definition of the currents (\ref{curr2}) is modified as 
\be
\Omega_\mu(x) = J_\mu^A(x) G^A
= e_\mu^a(x) P_a + K_\mu^{ab}(x) M_{ab} + B_\mu^{abc}(x) S_{abc}
\ee
where $e_\mu^a$ is 
regarded as the elfbein in constructing the gravity equation,
and the commutation relations for these fields/currents are given as 
\be
\left[K_0^{ab}(x), e_\mu^c(y)\right]\big|_{x_0=y_0}
&=& \left(\eta^{cb}e_\mu^a -\eta^{ca}e_\mu^b\right) \delta(\vx-\vy) \nt
\left[K_m^{ab}(x), e_n^c(y)\right]\big|_{x_0=y_0} &=& 0 \nt
\left[B_\mu^{abc}(x), e_\nu^d(y)\right]\big|_{x_0=y_0} &=& 0 \,.
\ee
In writing the commutators among elfbein $e_\mu^a$ (or translation $P_a$),
there are two choices:
{using the Poincar\'e algebra $SO(10,1)\times T_{11}$ (translation) or 
the $SO(10,2)$ algebra.}
Let us discuss them in the next subsections \ref{sec:poincare} and \ref{sec:10,2},
respectively.

\subsection{Inclusion of Poincar\'{e} algebra}
\label{sec:poincare}

One way {to define the commutation relations among elfbein} is as follows:
\be\label{eq43}
\left[e_0^a(x), e_0^b(y)\right]\big|_{x_0=y_0} &=& 0 \nt
\left[e_0^a(x), e_m^b(y)\right]\big|_{x_0=y_0} &=& iC\eta^{ab}\partial_m \delta(\vx-\vy) \nt
\left[e_m^a(x), e_n^b(y)\right]\big|_{x_0=y_0} &=& 0\,.
\ee
This comes from {the energy momentum tensor in the tangent space}
$P^a := \int e_0^a(x) d^{10}x$ with
\be
\left[P^a,P^b\right]=0\,.
\ee
By introducing the $\ell(1)$ vector representation operator $P^a$,
we can 
extend the $SO(10,1)$ group  to the eleven dimensional Poincar\'{e} group.
 This is the version we used in Refs.\,\cite{ref3,ref3a} but it has a disadvantage:
The resulting gravity equation deviates from the Einstein equation in a serious way. 

\subsection{Inclusion of $SO(10,2)$ algebra}
\label{sec:10,2}

We can correct this {problem} 
by assuming rather than the extension to the Poincar\'e group but to the {Anti-}de Sitter group $SO(10,2)$.
{This means that the $(10,1)$ translation $P^a$ is identified with $M^{a,11}$ in the $SO(10,2)$ algebra. Then}
we find that, combining 
the $SO(10,1)$ spin connection operator $\omega_\mu^{ab}$ which belongs to the $E(11)$ regular representation with 
the ``vector" representation operator corresponding to the tetrad field 
{$e_\mu^a \sim \omega_\mu^{a,11}$,}
they constitute the spin connection of $SO(10,2)$ algebra.
Thus we can again regard the tetrad as a part of the spin connection,
just as in Sec.\,\ref{sec:2}.
Then our gravity equation becomes the Anti-de Sitter gravity equation,
which was worked out in the $SO(3,2)$ case by Stelle and West~\cite{ref7} 
as shown below. 

Incidentally, this may imply
that we can go from $E(11)$ to $E(12)$, extending the M-theory to the F-theory.
This {should} happen because $SO(10,1)$ is a subgroup of $E(11)$ but $SO(10,2)$ is not.
{The regular representation of $E(12)$ should have the $SO(10,2)$ 
transformation $M$,
which includes both $M_{ab}$ in the $E(11)$ regular representation and $P_a$ in the $\ell(1)$ 
representation. Then we can expect that}
the regular representation of $E(11)$ combined with the $\ell (1)$ representation
would form the regular representation of $E(12)$.\footnote{
\begin{spacing}{1.0} \vspace{-14pt} \noindent
This needs to be studied more in detail in future works.
In Ref.\,\cite{ref3a}, we regard some operators in the $\ell(1)$ representation 
as the degrees of freedom of M-branes,
which correspond to the 2-form and 5-form central charges $Z$ in the $(10,1)$ supersymmetry algebra
$\{Q,Q\}\sim P+Z_2^M+Z_5^M$.
In the $(10,2)$ case, the supersymmetry algebra is $\{Q,Q\}\sim Z_2^F+Z_6^{F+}$.
Some researchers propose that $Z_2^F$ can be identified with the $SO(10,2)$ transformation $M_{ab}$~\cite{ref1_F,ref2_F},
and our discussion seems also along this approach.
The remaining $Z_6^{F+}$ should be a self-dual 6-form central charge 
describing the degrees of freedom of the brane expanding 
in $(6+2)$-dimensional space-time~\cite{ref2_F},
which is closely related to the $Z_5^M$ above. 
Therefore, at least, it is necessary to check whether this $Z_6^{F+}$ is also included in the $E(12)$ regular representation.
\end{spacing}
}



{For this purpose,}
all we need to do here 
is to change the commutation relation (\ref{eq43}) as
\be
\left[e_0^a(x), e_0^b(y)\right]\big|_{x_0=y_0} &=& -K_0^{ab}(x) \delta(\vx-\vy)  \nt
\left[e_0^a(x), e_m^b(y)\right]\big|_{x_0=y_0} &=& -K_m^{ab}(x) \delta(\vx-\vy) + iC\delta^{ab}\partial_m \delta(\vx-\vy) \nt
\left[e_m^a(x), e_n^b(y)\right]\big|_{x_0=y_0} &=& 0\,.
\ee
We now write down explicitly the equations for $e_\mu^a, \omega_\mu^{ab}$ and $B_\mu^{abc}$, following the recipe of the current algebra described above.
We start with the elfbein field/current and obtain
\be
D_{\mu a}^c e_\nu^a - D_{\nu a}^c e_\mu^a = 0
\ee
where we define the spin connection 
\be
D_{\mu a}^c = \delta_a^c \partial_\mu + \frac{i}{C}K_{\mu a}^c
=: \delta_a^c \partial_\mu + \omega_{\mu a}^c\,.
\ee
We also get the gauge fixing equation which we ignore in order to allow any other fixings:
\be
\partial_0 e_0^a = \partial_m e_m^a\,.
\ee
This conservation equation can be regarded as the gauge condition (Lorentz gauge) for $e_\mu^a$ which becomes a part of the $SO(10,2)$ gauge field as 
shown below. 

Next we consider the equation for $K_\mu^{ab}$ which is the same as in Refs.\,\cite{ref3,ref3a}.
Using the {definition of } 
$\omega_\mu^{ab}(x) = \frac{i}{C}K_\mu^{ab}(x)$, we have
\be
D_\mu^{cd,ab} \omega_{\nu ab} - D_\nu^{cd,ab} \omega_{\mu ab} 
&=& -\frac{1}{C^2}\left(e_\mu^c e_\nu^d -  e_\nu^c e_\mu^d\right)
+ \frac{1}{2C^2} \eta^{b[c} \left(B_\nu^{d]ea} B_{\mu abe} - B_\mu^{d]ea} B_{\nu abe}\right)
\ee
where
$
D_\mu^{cd,ab} = \eta^{ca}\eta^{db} \partial_\mu
 + \frac{1}{2}\left( \eta^{bc}\omega_\mu^{da} - \eta^{da}\omega_\mu^{bc}\right).
$

For the matter field $B_\mu^{abc}$, we obtain the equation as 
\be\label{eq54}
D_{\mu d}^{[c} B_\nu^{ab]d} - D_{\nu d}^{[c} B_\mu^{ab]d} = 0
\ee
where
$D_{\mu a}^c = \delta^c_a\partial_\mu + \omega_{\mu a}^c$.

We define the Riemann tensor as in Eq.(\ref{eq31}), 
then the gravity equation is written as
\be\label{eq56}
R^{cd}{}_{\mu\nu} 
&=& D_\mu^{cd,ab}\omega_{\nu ab} - D_\nu^{cd,ab}\omega_{\mu ab} \nt
&=& -\frac{1}{C^2}\left(e_\mu^c e_\nu^d - e_\nu^c e_\mu^d \right)
+ \frac{1}{2C^2} \eta^{b[c} \left(B_\nu^{d]ea} B_{\mu abe} - B_\mu^{d]ea} B_{\nu abe}\right) .
\ee
If we ignore the ``matter part'' $B$, 
this Eq.\,(\ref{eq56}) becomes identical to the Stelle-West's maximally symmetric solution~\cite{ref7} derived from their basically non-polynomial Lagrangian for the case of $SO(3,2)$. Our case is for the $SO(10,2)$ and we derived this equation from the current algebra. 

Let us here 
work backward (compared to Stelle-West~\cite{ref7}) and define the spin connection for $SO(10,2)$ in terms of tetrad
\be\label{eq:omega11}
\omega_\mu^{a,11}(x) := \frac{1}{C} e_\mu^a(x).
\ee
Then we have
\be
D_\mu^{c,11,ab}\omega_{\nu ab} - D_\nu^{c,11,ab}\omega_{\mu ab}
&=& \left[\eta^{ca}\eta^{11,b}\partial_\mu + \frac12 \left(\eta^{bc}\omega_\mu^{11,a}-\eta^{11,a}\omega_\mu^{bc}\right)\right]\omega_{\nu ab} - (\mu\leftrightarrow\nu) \nt
&=& D_{\mu a}^c e_\nu^a - D_{\nu a }^c e_\mu^a = 0
\ee
where $D_{\mu a}^c= \delta_a^c\partial_\mu + \omega^c_{\mu a}$.
If there is no $x^{11}$ dependence of $e_\mu^a(x)$, 
this is consistent with the following pure $(10,2)$ gravity (with no vector representations): 
\be
D^{cd,ab}_\mu \omega_{\nu ab} - D^{cd,ab}_\nu \omega_{\mu ab} = 0
\ee
where $\mu,\nu=0,1,\ldots,11$ for the curved space-time and 
$a,b,\ldots = 0,1,\ldots,11$ for the local flat space-time.

\subsection{``Matter'' field/current $B_\mu^{abc}$}

{In the above discussion we can reproduce the pure gravity} 
from the $E(11)$ viewpoint,
{but actually,}
it contains not just spin connection and tetrad 
but three-form 
antisymmetric tensor $B_\mu^{abc}$ within the gravity multiplet. 
From the $SO(10,1)$ viewpoint, this can be treated as a kind of matter field.

We note that the factor $1/C^2$ on the right hand side of Eq.\,(\ref{eq56})
can be completely absorbed {by 
using $\omega_\mu^{a,11}=\frac{1}{C}e_\mu^a$ instead of $e_\mu^a$ and 
by redefinition} 
\be
\text{new }B_\mu^{abc} := \frac{1}{C} \left(\text{old } B_\mu^{abc} \right).
\ee
This is expected since $\omega_\mu^{ab}$ and $B_\mu^{abc}$ belong to the same regular representation and it is known 
that these gauge field equations {should} have no dimensionful 
constants {in M-theory.} 

Of course, this does not mean the dimensionful constants disappear from our quantum theory. In the commutation relations among new $B$'s, 
{the constant $C$ still appears:}
\be
\left[B_0^{abc}(x),B_{0,def}(y)\right]\big|_{x_0=y_0}
&=&\frac{36i}{C}\delta^{[a}_{[d}\delta^b_e\omega_{0,f]}^{c]}\delta(\vx-\vy) \nt
\left[B_0^{abc}(x),B_{m,def}(y)\right]\big|_{x_0=y_0}
&=&\frac{36i}{C}\delta^{[a}_{[d}\delta^b_e\omega_{m,f]}^{c]}\delta(\vx-\vy)
+ \frac{6i}{C}\delta^{a}_{[d}\delta^b_e\delta^{c}_{f]}\partial_m \delta(\vx-\vy) \nt
\left[B_m^{abc}(x),B_{n,def}(y)\right]\big|_{x_0=y_0}
&=& 0\,.
\ee

\section{Negative cosmological constant for compactified space 
and positive one for our (3,1) space-time}

We learned above that adding $e_\mu^a$ which belong to the $E(11)$ vector representation
$\ell(1)$ 
amounts to enlarging $SO(10,1)$ to $SO(10,2)$,
and this gives 
the eleven dimensional Anti-de Sitter space
with the negative cosmological constant. 
{As we will see,}
this {provides} 
us a clue to obtain the de Sitter space in lower dimension in our formalism. 
We start with the current algebra formalism {with the} vector representation $\ell(1)$. 
The equation of motion (\ref{eq56}) can be rewritten 
using the new $B_\mu^{abc}$ and 
\be
\text{new } e_\mu^a := \frac{l_P}{C}(\text{old } e_\mu^a)
\ee
as
\be \label{eq63}
R^{cd}{}_{\mu\nu} 
&=& D_\mu^{cd,ab}\omega_{\nu ab} - D_\nu^{cd,ab}\omega_{\mu ab} \nt
&=& \frac{1}{l_P^2}\left(e_\mu^c e_\nu^d -  e_\nu^c e_\mu^d\right)
+\frac12\eta^{b[c} \left(B_\nu^{d]ea}B_{\mu abe} - B_\mu^{d]ea}B_{\nu abe}\right)
\ee
where $a,b,\ldots = 0,1,\ldots,10$.
Both $\omega_\mu^{ab}$ and $B_\mu^{abc}$ have the dimension of $\text{(length)}^{-1}$,
and $e_\mu^a$ is dimensionless.

Let us now compactify the directions $a,\mu = 4,\ldots,9$ and,
as in Eq.\,(\ref{eq:omega11}), define
\be
e_\mu^a(x) := C \omega_\mu^{a,11}(x)
\ee
in this compactified space.
Note that Eq.\,(\ref{eq63}) guarantees the Anti-de Sitter character of the compactified space but not the $(3,1)$ space-time.
Then we define the vielbein by 
\be\label{eq:omega10}
e_\mu^a(x) := \phi(x)^{-1} \omega_\mu^{a,10}(x)
\ee
for $a,\mu=0,1,2,3$.
The point is that $\phi(x)$ depends on space-time coordinate $x$ unlike the case of Stelle-West~\cite{ref7}.  
If $\phi(x)$ is chosen to be constant,
since we have a positive metric for the direction $10$, 
{the $(3,1)$ space-time becomes} 
de Sitter space rather than Anti-de Sitter space. 

From now on, we use the notation such that
\begin{itemize}
\item
the indices $a,b,c,\ldots$ (flat space) and $\mu,\nu, \rho,\ldots$ (curved space) 
only for the non-compact space-time $(0,1,2,3)$
\item
the indices $A,B,C,\ldots$ (flat space) and $M,N,L,\ldots$ (curved space) for the compactified space $(4,5,6,7,8,9)$
\item
{
the indices $\Gamma, \Delta, \Theta, \ldots$ (flat space) for all the directions 
in M-theory $(0,1,2,\ldots,10)$.}
\end{itemize}
In this notation, we have
\be\label{eq65}
e_\mu^a(x) &=& \phi(x)^{-1}\omega_\mu^{a,10}(x)
\qquad\text{for}~ a,\mu=0,1,2,3 \nt
e_M^A(x) &=& C\omega_M^{A,11}(x)
\qquad\text{for}~ A,M=4,5,\ldots,9 \nt
e_\mu^A(x) &=& e_M^a(x) = 0 
\qquad\text{otherwise}
\ee
and we define
\be
g_{\mu\nu}(x) = \eta_{ab}e_\mu^a(x)e_\nu^b(x)\,,\quad
g_{MN}(x) = \delta_{AB}e_M^A(x)e_N^B(x)\,.
\ee
Since the $\omega_\mu^{a,10}$ equation contains the contribution from matter field $B_\mu^{abc}(x)$, the $\phi(x)$ usually depends on space-time coordinate $x$. 
We will discuss this later in this paper.

{As a part of the gravity equation (\ref{eq56}),}
we have
\be\label{eq69}
R^{cd}{}_{\mu\nu}
&=& \partial_\mu\omega_\nu^{cd} - \partial_\nu\omega_\mu^{cd}
 - \omega_\mu^{ad}\omega_{\nu a}^c + \omega_\nu^{ad}\omega_{\mu a}^c
 - \omega_\mu^{Ad}\omega_{\nu A}^c + \omega_\nu^{Ad}\omega_{\mu A}^c \nt
&=&
 \phi^2\left(e_\mu^c e_\nu^d - e_\nu^c e_\mu^d\right)
 + \frac12\eta^{\Gamma[c} \left(B_\nu^{d]\Delta\Theta}B_{\mu \Theta\Gamma\Delta} - B_\mu^{d]\Delta\Theta}B_{\nu \Theta\Gamma\Delta}\right).
\ee
This is the equation for the $(3, 1)$ space-time with positive cosmological constant (or function) $\phi^2$.
We can also write down equations for other components. For example, the curvature for the compactified space satisfies
\be\label{eq70}
&&
R^{CD}{}_{MN}
 + \omega_M^{aD}\omega_{Na}^C - \omega_N^{aD}\omega_{Ma}^C
\nt &&\quad =
 -\kappa^2\left(e_M^C e_N^D - e_N^C e_M^D\right)
 + \frac12\eta^{\Gamma[C} \left(B_N^{D]\Delta\Theta}B_{M \Theta\Gamma\Delta}
 - B_M^{D]\Delta\Theta}B_{N \Theta\Gamma\Delta}\right).
\ee
This is the equation for the compactified space with negative cosmological constant. 
The equation for compactified space tetrad is
\be
D_{MC}^A e_N^C - D_{NC}^A e_M^C = 0\,.
\ee
The equation for the field $\phi(x)$ is obtained from the equation for $\omega_\mu^{a,10}(x)$ by substituting Eq.\,(\ref{eq65}), and we obtain
\be\label{eq72}
\partial_\mu\left(\phi e_\nu^c\right) - \partial_\nu\left(\phi e_\mu^c\right)
-\phi e_\mu^a\omega^c_{\nu a} + \phi\omega_\mu^{ca}e_{\nu a}
= \frac12\eta^{\Gamma[c} \left(B_\nu^{10]\Delta\Theta}B_{\mu \Theta\Gamma\Delta} - B_\mu^{10]\Delta\Theta}B_{\nu \Theta\Gamma\Delta}\right).
\ee
Assuming 
$\partial_\mu e^c_\nu - \partial_\nu e^c_\mu - e_\mu^a\omega^c_{\nu a} 
+ \omega_\mu^{ca}e_{\nu a} =0$,
which is needed to ensure the geometric interpretation explained above,
we have 
\be
\left(\partial_\mu\phi\right) \delta_\nu^\rho - \left(\partial_\nu\phi\right) \delta_\mu^\rho
= \frac12e^\rho_c\eta^{\Gamma[c} \left(B_\nu^{10]\Delta\Theta}B_{\mu \Theta\Gamma\Delta} - B_\mu^{10]\Delta\Theta}B_{\nu \Theta\Gamma\Delta}\right)
\ee
where $\Gamma,\Delta,\Theta$ covers 0 to 9,
since the direction $10$ gives vanishing contribution.
Suppose $\phi(x)$ depends only on time, $\phi=\phi(t)$, 
we obtain
\be \label{eq72,a}
\left(\partial_0\phi\right) \delta_j^k
= \frac12e^k_c\eta^{\Gamma[c} \left(B_j^{10]\Delta\Theta}B_{0 \Theta\Gamma\Delta}
 - B_0^{10]\Delta\Theta}B_{j \Theta\Gamma\Delta}\right)
\ee
and 
\be\label{eq72,b}
e^k_c\eta^{\Gamma[c} \left(B_j^{10]\Delta\Theta}B_{i \Theta\Gamma\Delta} - B_i^{10]\Delta\Theta}B_{j \Theta\Gamma\Delta}\right) = 0
\ee
where 
{$i,j,k,\ldots$ are spatial components $(1,2,3)$ in the curved spacetime.}
Thus the right hand side of Eq.\,(\ref{eq72,a})  must be proportional to $\delta^k_j$. 
The equation for $\phi$ will be discussed in Sec.\,\ref{sec:5} 
and it is crucial in the cosmological application.
Equations for the other curvature components can also  be written down  easily but we skip them.

\subsection{Further extension of the definition of $\phi$ field}

Equation 
(\ref{eq:omega10}) can be applied to all the internal components $\hat A=4,5,\ldots,10$
not just for the direction 10, 
defining the $e_\mu^a(x)$ current ``collectively'':
\be
\omega_\mu^{a,\hat A} = \phi_{\hat A} e_\mu^a
\ee
for $a,\mu=0,1,2,3$ and $\hat A=4,5,\ldots,10$.
Then Eq.\,(\ref{eq69}) becomes
\be\label{eq69+}
R^{cd}{}_{\mu\nu}
&=&
 \Bigl(\sum_{\hat A}\phi_{\hat A}^2\Bigr)\left(e_\mu^c e_\nu^d - e_\nu^c e_\mu^d\right)
 + \frac12\eta^{\Gamma[c} \left(B_\nu^{d]\Delta\Theta}B_{\mu \Theta\Gamma\Delta} - B_\mu^{d]\Delta\Theta}B_{\nu \Theta\Gamma\Delta}\right).
\ee
We may call this the ``vacuum democracy'' where all the components contribute 
{equally} to the vacuum energy.
Equations (\ref{eq72,a}) and (\ref{eq72,b}) become
\be \label{eq72,a'}
\left(\partial_0\phi_{\hat A}\right) \delta_j^k
= \frac12e^k_c\eta^{\Gamma[c} \left(B_j^{\hat A]\Delta\Theta}B_{0 \Theta\Gamma\Delta}
 - B_0^{\hat A]\Delta\Theta}B_{j \Theta\Gamma\Delta}\right)
\ee
and
\be \label{eq72a+}
e^k_c\eta^{\Gamma[c} \left(B_j^{\hat A]\Delta\Theta}B_{i \Theta\Gamma\Delta} - B_i^{\hat A]\Delta\Theta}B_{j \Theta\Gamma\Delta}\right) = 0\,.
\ee
{These equations must be solved,} combined with the equation for $B$:
\be
\partial_\mu B_\nu^{\Theta\Gamma\Delta} - \partial_\nu B_\mu^{\Theta\Gamma\Delta}
= \omega_{\mu \Xi}^{[\Theta} B_\nu^{\Gamma\Delta]\Xi} 
- \omega_{\nu \Xi}^{[\Theta} B_\mu^{\Gamma\Delta]\Xi}.
\ee

\subsection{Equivalence principle}
\label{sec:equiv}

Before going to apply the above formulation to cosmology, we discuss how the equivalence principle can be satisfied in our theory. 
Unlike the general covariance which was {automatically reproduced}, the equivalence principle turns out to require some conditions on the matter field and it is possible that there may be a matter field which potentially violates the equivalence principle. 
Our original energy momentum tensor to obtain the equations of motion is
\be
\Theta_{\mu\nu} = \frac{1}{C}\left(J_\mu^A J_\nu^A - \frac12\eta_{\mu\nu}\eta^{\rho\sigma} J_\rho^A J_\sigma^A\right).
\ee
The equivalence principle says that this must be equal to the right hand side of the Einstein equation in the local Lorentz frame. 
In our case, the Einstein equation can be obtained by taking the trace of Eq.\,(\ref{eq56})
which reads (omitting the cosmological term and using the old $B$, only in this subsection)
\be
R_{\mu\nu} - \frac12 g_{\mu\nu}R
&=& \frac{1}{2C^2}e_c^\sigma e_{d\mu} \eta^{b[c} 
 \left(B_\nu^{d]ea}B_{\sigma abe} - B_\sigma^{d]ea}B_{\nu abe} \right)
\nt &&
-\frac{1}{4C^2}g_{\mu\nu} e_c^\sigma e_d^\rho \eta^{b[c} 
 \left(B_\rho^{d]ea}B_{\sigma abe} - B_\sigma^{d]ea}B_{\rho abe} \right)
\ee
where $a,b,\ldots$ and $\mu,\nu,\ldots$ run over $0,1,\ldots,10$
(only in this subsection).
This requires that 
\be\label{eq73}
\frac{1}{2l_P^2 C^2} e_c^\sigma e_{d\mu} \eta^{b[c} 
 \left(B_\nu^{d]ea}B_{\sigma abe} - B_\sigma^{d]ea}B_{\nu abe} \right)
\stackrel{!}{=} \frac{V}{C}B_{\mu abe}B_\nu^{abe}
\ee
where $V$ is the volume of the compactified space.
If we assume $B_{abcd} := e_a^\mu B_{\mu bcd}$ is a totally antisymmetric tensor,
meaning that the product representation $e_a^\mu\otimes B_{\mu bcd}$
must belong to an irreducible representation of $SO(10,1)$,
then the left hand side of Eq.\,(\ref{eq73}) becomes
\be\label{eq73+}
\frac{1}{2l_P^2 C^2} e_c^\sigma e_{d\mu} \eta^{b[c} 
 \left(B_\nu^{d]ea}B_{\sigma abe} - B_\sigma^{d]ea}B_{\nu abe} \right)
= \frac{1}{2l_P^2C^2}B_{\mu abe}B_\nu^{abe}.
\ee
This shows that the equivalence principle is satisfied under the assumption for
$B_{abcd}$ with
\be
2 l_P^2 C = \frac{1}{V}\,.
\ee

\section{Application to cosmology}
\label{sec:5}

First, we assume that the classical gravity is described by Einstein equation rather than our quantum equation (\ref{eq63}) which leads to the Einstein equation by contracting indices.
Precisely what this means is the following.

We define the classical metric as
\be
g_{\mu\nu} = \langle e_\mu^a e_{\nu a}\rangle
= \sum_n \langle c |e_\mu^a| n\rangle \langle n|e_{\nu a}|c\rangle
\ee
where $|c\rangle$ corresponds to a certain classical state and
$\sum_n |n\rangle\langle n|=1$.
The only case we can define the classical $e_\mu^a$ is when $\sum_n |n\rangle\langle n|$
can be replaced by $|c\rangle\langle c|$, and it is not generally correct.
{Instead,} we can define the classical $e_\mu^a$ by $g_{\mu\nu} = e_\mu^a e_{\nu a}$
when $g_{\mu\nu}$ is given by solving the Einstein equation,
but this $e_\mu^a$ would not satisfy our original equation (\ref{eq63}). 
The similar argument can be done for $\omega_\mu^{cd}(x)$.

The Einstein equation can be derived by taking the trace of Eq.\,(\ref{eq69+})
and it reads
\be
R_{\sigma\nu} 
= 3\Bigl(\sum_{\hat A}\phi_{\hat A}^2\Bigr)g_{\sigma\nu}(x)
+ \frac12 e_c^\mu e_{d\sigma} \eta^{\Gamma[c} \left(B_\nu^{d]\Delta\Theta}B_{\mu \Theta\Gamma\Delta}
- B_\mu^{d]\Delta\Theta}B_{\nu \Theta\Gamma\Delta} \right).
\ee
$R_{\mu\nu}$ can be expressed in terms of geometrical variable 
$g_{\mu\nu}$ rather than the spin connection as explained in Sec.\,\ref{sec:2}.
{Let us solve} 
this equation,
combining it with the equations for $\phi_\hA$ (\ref{eq72}) and 
for $B_\mu^{ade}$ (\ref{eq54}).

We now interpret this system of the equations  as a classical system and apply it to the homogeneous and isotropic universe where the energy momentum tensor must take the form of perfect liquid:
\be
B_\mu^{\Delta\Gamma\Theta}B_{\nu \Delta\Gamma\Theta}
-\frac12 g_{\mu\nu} B^{\sigma \Delta\Gamma\Theta}B_{\sigma \Delta\Gamma\Theta}
= pg_{\mu\nu} + (p+\rho) U_\mu U_\nu
\ee
where
$U_\mu = (1,0,0,0)$, 
$g_{\mu\nu} = {\rm diag. }(-1,g_{ij})$.
Then we obtain
\be
\rho &=& \Theta_{00} = B_0^{\Delta\Gamma\Theta} B_{0\Delta\Gamma\Theta} + \frac12 B^{\sigma \Delta\Gamma\Theta} B_{\sigma \Delta\Gamma\Theta}
\nt
p g_{ij} &=& \Theta_{ij} = B_i^{\Delta\Gamma\Theta} B_{j \Delta\Gamma\Theta} - \frac12 g_{ij} B^{\sigma \Delta\Gamma\Theta} B_{\sigma \Delta\Gamma\Theta}.
\label{eq81}
\ee
Using the equivalence condition 
$B_{\mu \Delta\Gamma\Theta} = e_\mu^d B_{d\Delta\Gamma\Theta}$,
we can rewrite it as
\be
p g_{ij} = e_{id} B^{d\Delta\Gamma\Theta} e_j^e B_{e\Delta\Gamma\Theta} - \frac12 g_{ij} B^{{d} \Delta\Gamma\Theta} B_{{d} \Delta\Gamma\Theta}.
\ee
Introducing $p^{(1)}$ and $p^{(2)}$ by
\be
B^{d\Delta\Gamma\Theta} B_{e\Delta\Gamma\Theta} = p^{(1)} \delta^d_e + p^{(2)} U^d U_e\,,
\ee
we obtain
\be
\rho = p^{(1)} + \frac12 p^{(2)},\quad
p = -p^{(1)} + \frac12 p^{(2)}.
\ee
Adding the $\phi_\hA$ term, we have
\be
\rho_\text{tot} = \rho + \rho_\phi\,,\quad
p_\text{tot} = p+p_\phi\,; \quad
\rho_\phi = -p_\phi = 6\sum_\hA \phi_\hA^2\,.
\ee
Then the condition of accelerating universe
$\rho_\text{tot} + 3p_\text{tot} < 0$ becomes
\be\label{cond:acc}
-6\sum_\hA \phi_\hA^2 - p^{(1)} + p^{(2)} < 0\,.
\ee 

\subsection{Simple solution}

We now obtain a simple solution which automatically satisfies,
1. perfect fluid condition, 
2. accelerating condition (\ref{cond:acc}), 
3. Inflation-like property at the initial universe and 
4. presumably similar to quintessence model at the later stage of the universe.
We will also discuss the equivalence principle in the next subsection.

From now on, we assume 
the directions $i,j,\cdots =1,2,3$ correspond to $r,\theta,\varphi$ 
(polar coordinates)
and $e_i^a$ is diagonal as 
the homogeneous 
and isotropic 
universe requires.
In Eq.\,(\ref{eq72,a'}), to make the right hand side proportional to 
$\delta_j^k$, 
{we impose here}
\be
e^{\Gamma k}\left( B_j^{\hA\Delta\Theta}B_{0,\Theta\Gamma\Delta} 
- B_0^{\hA\Delta\Theta}B_{j,\Theta\Gamma\Delta} \right)
&\propto& \delta^k_j \nt
e_a^k\eta^{\Gamma\hA}\left( B_j^{a\Delta\Theta}B_{0,\Theta\Gamma\Delta} 
- B_0^{a\Delta\Theta}B_{j,\Theta\Gamma\Delta} \right)
&\propto& \delta^k_j,
\ee
{separately.}
Then we must have
\be\label{eq86}
B_j^{\hA\Delta\Theta}B_{0,\Theta\Gamma\Delta} 
- B_0^{\hA\Delta\Theta}B_{j,\Theta\Gamma\Delta}
&=:& e_{j \Gamma} F_\hA(t) \nt
B_j^{a\Delta\Theta}B_{0,\Theta\Gamma\Delta} 
- B_0^{a\Delta\Theta}B_{j,\Theta\Gamma\Delta}
&=:& -e_j^a \delta^\hA_\Gamma \tF_\hA(t)
\ee
for certain functions $F_\hA(t), \tF_\hA(t)$,
and Eq.\,(\ref{eq72,a'}) can be solved as
\be
\phi_\hA(t) = \frac14 \int_{t_0}^t \left(F_\hA(t) + \tF_\hA(t) \right) dt + \Omega_\hA
\ee
where $\Omega_\hA$ is an integral constant.
Note that Eq.\,(\ref{eq86}) with $\Gamma=0$ requires
\be\label{eq86a}
B_j^{\Xi\Delta\Theta}B_{0,\Theta 0\Delta} 
- B_0^{\Xi\Delta\Theta}B_{j,\Theta 0\Delta}  = 0\,,
\ee
then we set $B_\mu^{0\Delta\Theta}=0$.
{Moreover, in order to obtain a simple solution, we impose}
the following ansatz: 
\be
B_\mu^{\hA \alpha\beta}(x) &=:& \epsilon^{\alpha\beta\gamma} G^{\hA}_{\mu \gamma}(x) \nt
B_{\mu}^{\alpha\beta\gamma}(x) &=:& \epsilon^{\alpha\beta\gamma} H_\mu(x)
\label{eq91}
\ee
{where $\alpha,\beta,\ldots=1,2,3$ in the flat space,
and all the other components of $B_\mu^{\Xi\Delta\Theta}=0$.}
Using this ansatz, 
Eq.\,(\ref{eq86}) becomes
\be\label{eq86+}
e_{j\beta} F_\hA(t)
&=& B_j^{\hA \gamma\alpha}B_{0,\alpha\beta\gamma} 
- B_0^{\hA \gamma\alpha}B_{j,\alpha\beta\gamma}
= 2G^\hA_{j\beta} H_0 - 2G^\hA_{0\beta} H_j \nt
e_{j\beta} \tF_\hA(t)
&=& -B_{j,\beta\alpha\gamma} B_0^{\hA\alpha\gamma} 
+ B_{0,\beta\alpha\gamma} B_j^{\hA\alpha\gamma}
= -2G^\hA_{0\beta} H_j + 2G^\hA_{j\beta} H_0 \,.
\ee
We now solve the equation for $B_\mu^{abc}$ field/current (\ref{eq54}):
\be \label{eq54+}
\partial_\mu B_\nu^{abd} - \partial_\nu B_\mu^{abd}
&=& 
-\left(\omega_{\mu e}^{[d} B_\nu^{ab]e} - \omega_{\nu e}^{[d} B_\mu^{ab]e}\right)
-\phi_\hA \left(e_{\mu}^{[d} B_\nu^{ab]\hA} - e_{\nu}^{[d} B_\mu^{ab]\hA}\right)
\nt
\partial_\mu B_\nu^{\hA bd} - \partial_\nu B_\mu^{\hA bd}
&=& 
-\left(\omega_{\mu e}^{[d} B_\nu^{\hA b]e} - \omega_{\nu e}^{[d} B_\mu^{\hA b]e}\right)
-\phi_\hB \left(e_{\mu}^{[d} B_\nu^{\hA b]\hB} - e_{\nu}^{[d} B_\mu^{\hA b]\hB}\right),
\ee
and obtain, assuming $B_\mu^{\hA \hB c}=0$ for consistency,
\be
D_{\mu a}^\beta G^\hA_{\nu \beta} -  D_{\nu a}^\beta G^\hA_{\mu \beta} = 0\,,\quad
\partial_{\mu} H_{\nu} - \partial_{\mu} H_{\nu} = 0\,,
\ee
where $D_{\mu a}^\beta = \delta_a^\beta\partial_\mu + \omega_{\mu a}^\beta$.
Then we can define
\be
G^\hA_{\mu \beta}(x) =: e_{\mu \beta}(x) G_\hA\,,\quad
H_\mu(x) =: \partial_\mu H(x)\,,
\ee
and Eq.\,(\ref{eq86+}) becomes
\be
F_\hA(t) = \tF_\hA(t) = 2 G_\hA \partial_0 H(x)\,,
\ee
which tells us that $H(x)$ must be a function of time $t$,
{and then the condition (\ref{eq72a+}) is also automatically satisfied.}
Finally, we obtain
\be
\phi_\hA(t) 
= \frac14\int_{t_0}^t \left(F_\hA(t) + \tF_\hA(t) \right) dt + \Omega_\hA
= G_\hA\left(H(t)-H(t_0)\right) + \Omega_\hA\,.
\ee

{In order to determine 
the time dependence of $H(t)$, 
or equivalently that of $\phi_\hA(t)$,
we need to use} the 
Einstein equation 
\be
R_{\mu\nu} - \frac12 g_{\mu\nu} R
=
-3\Bigl(\sum_\hA \phi_\hA^2 + \sum_\hA G_\hA^2\Bigr) g_{\mu\nu}
+2\sum_\hA G_\hA^2 e_\mu^i e_{\nu i}
= \kappa \Theta_{\mu\nu}
\ee
with $\kappa = l_P^2$.
This shows that our solution satisfies the  perfect fluid condition with the following density and pressure
\be \label{eq104}
\rho &=& \Theta_{00} =  \frac{3}{\kappa}\left(\sum \phi_\hA^2+\sum G_\hA^2\right) \nt
p g_{ij} &=& \Theta_{ij} = -\frac{1}{\kappa}\left(3\sum \phi_\hA^2+\sum G_\hA^2\right)g_{ij}\,.
\ee
Note that
the accelerating condition is also automatically satisfied:
\be
\rho + 3p = -\frac{6}{\kappa}\sum \phi_\hA^2 < 0\,.
\ee
The point here is that $\sum G_\hA^2$ term satisfies $p=-\frac13\rho$ and does not contribute to the acceleration.
Coexisting with 
the $\sum \phi_\hA^2$ term which satisfies $p=-\rho$,
this term can be time independent 
as we now show. 

The Einstein equation for the Friedmann–Lema\^{i}tre–Robertson–Walker metric
(with the curvature $k=0$) 
\be\label{eq107}
\left(\frac{\partial_0 a}{a}\right)^2 = {\frac{\kappa}{3}}\rho\,,\quad
{\partial_0 \rho} + \frac{3\partial_0 a}{a}(\rho+p) = 0
\ee
is solved as
\be
a = a_0\exp\left[-\frac{\sum \phi_\hA^2}{2\sum G_\hA^2}\right]
\ee
and 
\be \label{eq112}
\sum \phi_\hA^2 = 
\left(\pm  \sum G_\hA^2 \cdot t + \Omega_0 \right)^2 - \sum G_\hA^2 \,.
\ee
where 
$\Omega_0$ is an integral constant.
{If this constant is negative for the $+$ sign or is positive for the $-$ sign,
at the early time $t\ll 1$,}
this is the same as the ``Chaotic Inflation"~\cite{ref8}.\footnote{
There are, as is well known, many other versions of inflation models starting from the original proposal by Refs.\,\cite{ref8a,ref8b,ref8c,ref8d}.
They are well summarized in, for example, Refs.\,\cite{ref8e,ref8f}.}

The most important and interesting aspect of this solution is that we can add  arbitrary constant vacuum energy (plus or minus) to the $\phi_\hA^2$ term
{without 
the consequence changed.} 
When we add arbitrary constant $\Lambda$ to the $\sum\phi_\hA^2$ term in Eq.\,(\ref{eq104}):
\be \label{eq104a}
\rho &=& \frac{3}{\kappa}\left(\sum \phi_\hA^2+\Lambda+\sum G_\hA^2\right) \nt
p g_{ij} &=& -\frac{1}{\kappa}\left(3\sum \phi_\hA^2+3\Lambda+\sum G_\hA^2\right)g_{ij}\,,
\ee
we obtain
\be \label{eq112+}
\sum \phi_\hA^2 = 
\left(\pm \sum G_\hA^2 \cdot t + \Omega_0 \right)^2 - \sum G_\hA^2 - \Lambda\,.
\ee
Note that we can always choose $\sum G_\hA^2+\Lambda > 0$.
Then 
the inflation factor $a$ 
starts with some initial value and ends up to be 
the maximum value 
at certain time $t=\frac{|\Omega_0|}{\sum G_\hA^2}$,
which would be identified as the end of inflation as we discuss later.
{This behavior is the same as Eq.\,(\ref{eq112})
especially in that the inflation factor $a$ is independent from $\Lambda$,
since a shift of $\Lambda$ can absorbed in the factor $a_0$.
At later time, only the quantum fluctuation of $\phi_\hA$ and $G_\hA$ 
can survive till even the current universe,
and would be 
a candidate for the quintessence~\cite{ref9a,ref9b,ref9c,ref9d,ref9e,ref9f,ref9g,ref9h,ref9i,ref9j}. 
This issue will be discussed shortly. 

The constant $\Lambda$ may arise from supersymmetry breaking, standard model symmetry breaking, or just from electron-positron vacuum energy. However, they 
contribute to the expansion of the universe neither at the initial stage nor at the current stage.

\subsection{Equivalence principle}
\label{sec:ep}

Equation (\ref{eq73+}) becomes
\be
e_c^\sigma e_{d\mu} \eta^{b[c} 
 \left(B_\nu^{d]\Delta\Theta}B_{\sigma \Theta\Gamma\Delta} - B_\sigma^{d]\Delta\Theta}B_{\nu \Theta\Gamma\Delta} \right)
= B_{\mu \Theta\Gamma\Delta}B_\nu^{\Theta\Gamma\Delta}.
\ee
The left hand side is calculated to be
$12\sum G_\hA^2e_\sigma^i e_{\nu i}$,
while the right hand side is given by
$2\sum G_\hA^2e_\sigma^i e_{\nu i} + 6\delta_\sigma^0\delta_\nu^0 (\partial_0 H)^2$.
This shows that the equivalence principle is not satisfied by our solution.
{However,} 
we don't regard this as the defect of our theory but something to be checked experimentally:
The field $H$ does not couple to gravity except through $\phi_\hA$ as shown in Eq.\,(\ref{eq54+}),
and $G^\hA_{\mu a}$ couples to gravity violating the equivalence principle by a factor of 6.

\subsection{History of universe after the inflation-like initial stage}
\label{sec:5C}

As we saw above the combination of two fields, 
$\phi_\hA$ which is the conformal field with $\rho+p=0$ and 
$G_\hA$ with $\rho+3p=0$, 
gives rise to the inflation-like behavior at the initial stage of the universe. 
The field $G_\hA$ is a component of the three-form 
antisymmetric tensor
$B_\mu^{abc}$ which appears in the M-theory  formulated as E(11) current-current theory~\cite{ref3,ref3a}. Due to 
the coexistence of these two fields, the usual 
$\rho\sim t^{-3(1+w)}$ ($w=p/\rho$) rule does not work 
and the equation of motion gives constant $G_\hA$ 
and the continuity equation gives $\rho_{\phi}$ to be quadratic in time.

At the end of inflation, $\sum\phi_\hA^2$ takes the minimum value and its time derivative vanishes. $G_\hA$ stays constant but it cannot stay by itself to satisfy the continuity equation. The reheating process would produce ordinary particles, and 
$\phi_\hA$ with constant $G_\hA$ would coexist with these ordinary particles. What are the ordinary particles? We customarily think of 
dark matter candidate, radiation and nuclei. 

Then the question we want to raise here is: Do we really need dark matter? Or rather, what is the dark matter?

Here we want to exploit the possibility that $G_\hA$ itself plays the role of dark matter. 
This possibility is implied by comparing the usual dark matter which has $p=0$ and
$G_\hA$ with $w=p/\rho=-1/3$. 
The quintessence is usually defined to be an object which has $w$ with $-1<w<0$, and so one might think of regarding $G_\hA$ with $w=-1/3$ as a candidate for the quintessence.
Actually $G_\hA$ can behave like the quintessence, however, interestingly,
it can also play the role of dark matter as we show below. 

When the inflation is terminated and reheating is starting, the universe may be described by two components, constant $G_\hA$ and ordinary relativistic particles with $p=\frac13 \rho$
(since we naively expect that $\phi_\hA$ keeps taking its minimum value after the inflation).
We assume the ordinary particle density and pressure are given by
\be
\rho=\frac{3}{\kappa} \xi\,,\quad
p=\frac{1}{\kappa} \xi\,,\quad
\xi>0
\ee
where $\xi$ is generally time dependent.

Then Einstein equation (\ref{eq107}) takes the form: 
\be
\left(\frac{\partial_0 a}{a}\right)^2 = \xi + \sum G_\hA^2\,,\quad
3\partial_0 \xi + \frac{3\partial_0 a}{a}\left(4\xi + 2\sum G_\hA^2\right) = 0\,,
\ee
and its solution is 
\be
a \sim \left(\xi+\frac12\sum G_\hA^2\right)^{-\frac14} 
= \left(\frac{2}{G^2}\sinh^2 (\sqrt{2G^2}\cdot t)\right)^{\frac14},
\ee
where $G^2:=\sum_\hA G_\hA^2$.
This is a very slow expansion compared to fast decrease in the particle density. Usually what is required for the expansion at this stage is to have a slow expansion to get  ample time for the inhomogeneity created by the quantum fluctuation to develop enough. We have successfully a very slow expansion here, but maybe too slow.

If we needs a faster expansion, we may add a portion of $\phi_\hA$ as
\be
\rho_\phi = \frac{3}{\kappa}\sum \phi_\hA^2 =: \alpha \rho\,,\quad
p_\phi = -\alpha\rho\,,
\ee
which means that 
a certain amount of the leftover $\phi_\hA$ remains in the reheating stage.
Then the equation becomes
\be
\left(\frac{\partial_0 a}{a}\right)^2 = (1+\alpha)\xi + \sum G_\hA^2\,,\quad
3(1+\alpha)\partial_0 \xi + \frac{3\partial_0 a}{a}\left(4\xi + 2\sum G_\hA^2\right) = 0\,,
\ee
and its solution is
\be
a\sim \left(\xi+\frac12\sum G_\hA^2\right)^{-\frac{1+\alpha}{4}}
= \left(\frac{2(1+\alpha)}{(1-\alpha)G^2}\sinh^2 (\frac{\sqrt{2(1-\alpha)G^2}}{1+\alpha}\cdot t)\right)^{\frac{1+\alpha}{4}}.
\ee
Depending on the value of $\alpha$, we may have an appropriate expansion speed of the universe with the limitation that  the clustering can develop enough. 

The role of clustering which is usually attributed to the dark matter is played by $G_\hA$ here. It has $w=p/\rho=-1/3$. Since it has 6 times stronger gravitational force implied by the equivalence principle (Sec.\,\ref{sec:ep}) and the negative pressure, it is very susceptible to gravitational collapse and 
is likely to exist in the form of black hole. All the clusterings, stars, galaxies and superstructures could contain the $G_\hA$ black holes in their center. Of course, it is possible that the WIMP dark matter 
forms a black hole with much smaller mass than the Chandrasekar or Oppenheimer-Volkov mass limit because of its small pressure, but in this case, a part of the dark matter would take the form of free particle and may be found by some method of dark matter search. In our case, since $G_\hA$ 
exists only in the form of black hole and a tiny mass black hole is possible,
any particle-like dark matter search would fail.

\subsection{Current Universe}

We have been discussing the early universe so far within our theoretical framework. 
Let us now discuss the current universe.

After the density of the ordinary particles become small
({more precisely,} 
smaller than the $\phi_\hA$ density which is a leftover from the inflational stage due to the quantum fluctuation), 
the field $\phi_\hA$ together with $G_\hA$ again takes over the ordinary particle density.
This $G_\hA$ is in fact a peculiar object which satisfies $\rho+3p=0$,
and would be a replacing dark matter.

Suppose the current universe is mostly made up of $\phi_\hA$ with density $\rho$ and $G_\hA$ 
with density $\alpha'\rho$.
Then the total density is $(1+\alpha')\rho$ and the total negative pressure is $(1+\frac{\alpha'}{3})\rho$.
Experimentally, we know that
$(\text{total density}) / (\text{total negative pressure}) \sim 1.5$,
then we obtain $\alpha'\sim 1$.
This means that about half of the cosmic density comes from $G_\hA$ 
with negative pressure and violating the equivalence principle;
the other half comes from $\phi_\hA$.
In this way, the combination of $\phi_\hA$ and $G_\hA$ would play the role of quintessence
in our current universe.

Several points have to be clarified to validate our assertions:

\begin{enumerate}
\item 
We are asserting that the combination of the two fields (currents) gives something similar to the chaotic inflation model. The two fields are specific components of the three-form 
tensor $B_\mu^{abc}$ which belongs to the same multiplet as gravity in $E(11)$ group. 
One field $\phi_\hA$ has $\rho+p=0$ just as the inflaton and behave as it.
The other field $G_\hA$ has negative pressure with $\rho+3p=0$, which is playing the role of dark matter in the matter dominant universe.
In the current universe, both fields play the role of quintessence.

\item
This result shows a very different picture of the universe from the regular inflation model: We do not need dark matter in the universe but it is replaced by the field $G_\hA$ with negative pressure with $\rho+3p=0$. The dark matter's role is to provide a sufficient clustering at the matter dominant stage. $G_\hA$ would be able to play this role, since it has 6 times larger attractive force than the regular matter with equivalence principle and it has negative pressure unlike 
the ordinary dark matter (massive weak interacting matter)
with vanishing pressure. 
Since $G_\hA$ has only gravitational interaction, the most likely form of its existence 
is 
a black hole or a heavy object like neutron star, when it coexists with ordinary particles. 
The ordinary neutron star has the mass range of $1.2-1.8$ solar mass, and a much lighter pulsar could also account for a star with our $G_\hA$ at its center. 
This 
speculation must be substantiated by actual calculations and experiments. 
As we mentioned at the end of Sec.\,\ref{sec:5C},
finding no particle-like dark matter but the black hole dark matter would be the first evidence of our approach. 
\end{enumerate}

\section{Discussion of our results and concluding remarks}

We have demonstrated that quantum gravity can be formulated using the current algebra scheme. Gauge principle is the basic principle: neither general covariance nor equivalence principle are used to formulate the gravity, but the former is automatically and the latter can be satisfied under certain conditions. This leaves a room for some matter to violate the equivalence principle, although we have no such experimental or observational  candidate at this time.
{Actually,} 
we presented here that what we usually call dark matter is a candidate to violate the equivalence principle.

The only basic gauge field for the gravity is the spin connection.  The tetrad can be defined from this field using the Stelle-West ansatz~\cite{ref7}, and then
all the other geometric objects like
$g_{\mu\nu}, \Gamma_{\mu\lambda}^\nu$ and $R^\lambda{}_{\sigma\mu\nu}$ 
are defined in terms of the spin connection gauge field. Our gravity equation is for all the components of $R^\lambda{}_{\sigma\mu\nu}$ and not just for the $R_{\mu\nu}$ as in Einstein equation. The latter can be obtained  by taking the trace of the former equation. 
The adoption and the extension of the Stelle-West ansatz make it possible to have a negative cosmological constant for the compactified space and a positive one for our $(3,1)$ space-time. 

We have applied the current algebra formalism to the $SO(N,1)$ and $SO(N-1,2)$ gauge groups, but the most interesting result is obtained when we apply this formalism to $E(11)$ Kac-Moody algebra. Gravity multiplet contains not just the spin connection $\omega_\mu^{ab}$ but the antisymmetric three-form 
tensor $B_\mu^{abc}$ 
from the viewpoint of $SO(10,1)$ subgroup of $E(11)$.  This $B_\mu^{abc}$ field can be regarded as a matter field from the $SO(10,1)$ viewpoint. In fact, the {corresponding} three-form 
antisymmetric tensor $C_{\mu\nu\rho}$ is the only bosonic field in low energy eleven dimensional supergravity Lagrangian. 

Since $E(8)$ is a subgroup of $E(11)$, our theory is $E(8)$ invariant as in the usual formulation of M-theory~\cite{ref10}. The $E(8)$ current algebra using only its regular representation is supposed to cover the entire hadrons and leptons. Its relation to gravitational multiplet considered in our study ($\omega_\mu^{ab}, e_\mu^a, B_\mu^{abc}$ and $\psi_\mu$) must be worked out. This is left for the future work.

The classical limit of our formalism was applied to cosmology with the time-dependent factor $\phi_\hA$ ($\hA=4,5,\ldots,10$) which appears in the Stelle-West ansatz~\cite{ref7} and the field $B_\mu^{abc}$. Then the inflationary expansion of the universe emerges as a simple solution. 
The $\phi_\hA$ field (together with some components of $B_\mu^{abc}$) plays the role of ``inflaton" and its time dependence is determined by energy-momentum conservation law which is a part of Einstein equation.  
$B_\mu^{\hA ab}$ ($\mu,a,b=0,1,2,3$) 
becomes a time independent perfect fluid with $p=-\frac13\rho$.
Its density is constant due to its coexistence 
with $\phi_\hA$,
as our solution shows. The time dependence of $\phi_\hA$ gives exactly the same behavior
as the inflating universe in the 
``chaotic inflation model"~\cite{ref8},
although the principle, the ingredient and the derivation are quite different. 

Moreover, 
we used the $\phi_\hA$ field and the $B_\mu^{\hA ab}$ components 
as a candidate for ``quintessence" \cite{ref9a,ref9b,ref9c,ref9d,ref9e,ref9f,ref9g,ref9h,ref9i,ref9j} which is to explain the accelerating universe at the present time. 
These fields also explain the behavior of the matter dominant universe as well as the current universe:
Especially 
the $B_\mu^{\hA ab}$ components behave like the dark matter,
and they are susceptible 
to gravitationally collapse and to form a black hole.
Then it would be related to the primordial black hole as the candidate of dark matter~\cite{ref11}.\footnote{
They refer to many other works done on this topic starting with such classic papers as Refs.\,\cite{ref11a,ref11b,ref11c}.}
We would like to clarify this relation in future works.

\subsection*{Acknowledments}

The authors would like to thank
Professors Satoshi Iso and Peter West for their interest in our work.

\begin{thebibliography}{99}
\bibitem{ref1}
R. Uchiyama, Phys. Rev. {\bf 101}, 1597 (1956). 
\bibitem{ref1a}
T. W. B. Kibble, J. Math. Phys. {\bf 2}, 212 (1961).
\bibitem{ref1b}
C. Moller, K. Dan. Vidensk. Selsk. Mat. Fys. Skr. 1 (No.10), 1 (1961).
\bibitem{ref1c}
C. Pellegrini and J. Plebanski, K. Dan. Vidensk. Selsk. Mat. Fys. Skr. 2 (No.2), 1 (1962).
\bibitem{ref1d}
K. Hayashi and T. Nakano, Prog. Theor. Phys. 38, 491 (1967).
\bibitem{ref1e}
F. W. Hehl, in Cosmology and Gravitation, ed. by P. G. Bergmann and V. de Sabbata (Plenum, New York, 1980).
\bibitem{ref1f}
F. Gronwald and F. W. Hehl, On the Gauge Aspects of Gravity, in Proceedings of the 14th School of Cosmology and Gravitation, Erice, Italy, ed. by P. G. Bergmann, V. de Sabbata and H.-J Treder (World Scientific, Singapore, 1996).
\bibitem{ref1g}
M. Blagojevi{\'c}, Gravitation and Gauge Symmetries (IOP Publishing, Bristol, 2002).

\bibitem{ref2}
P. West,  Class. Quant. Grav. {\bf 18} (2001) 4443 [arXiv:hep-th/0104081].
\bibitem{ref2a}
P. West, Phys. Lett. {\bf B575} (2003) 333-342 [arXiv:hep-th/0307098].
\bibitem{ref2b}
A. Tumanov and P. West,  Phys. Lett. {\bf B759}  (2016) 663 [arXiv:1512.01644 [hep-th]].
\bibitem{ref2c}
A. Tumanov and P. West, Phys. Lett. {\bf B758} (2016) 278 [arXiv:1601.03974 [hep-th]].
\bibitem{ref2d}
P. West, ``A brief review of $E$ theory,'' Proceedings of Abdus Salam's 90th  Birthday meeting, 25-28 January 2016, NTU, Singapore,  Editors L. Brink, M. Duff and K. Phua, World Scientific Publishing and IJMPA,  Vol 31, No 26 (2016) 1630043. 

\bibitem{ref3}
H. Sugawara, Int. J. Mod. Phys. A {\bf 32} (2017) no.05, 1750024 [arXiv:1701.06894 [hep-th]].
\bibitem{ref3a}
S. Shiba and H. Sugawara, Int. J. Mod. Phys. A {\bf 33} (2018) no. 07, 1850051
[arXiv:1709.07169 [hep-th]].

\bibitem{ref4}
K. Bardakci, Y. Frishman and M. B. Halpern, Phys. Rev. {\bf 170} (1968) 1353.

\bibitem{ref5}
H. Sugawara, Phys. Rev. {\bf 170} (1968) 1659.

\bibitem{ref6}
J. Schwinger, Phys. Rev. {\bf 130} (1963) 406; ibid. {\bf 130} (1963) 800.

\bibitem{ref7}
K. S. Stelle and P. C. West,  Phys. Rev. D {\bf 21}, 1466 (1980). 

\bibitem{ref1_F}
S. Hewson and M. Perry,
Nucl. Phys. B {\bf 492}, 249-277 (1997)
[arXiv: hep-th/9612008].
\bibitem{ref2_F}
S. F. Hewson,
Nucl. Phys. B {\bf 534}, 513-530 (1998) 
[arXiv:hep-th/9712017].

\bibitem{ref8}
A. Linde, Phys. Lett. B {\bf 129}, 177 (1983).

\bibitem{ref8a}
A. H. Guth, Phys. Rev. D {\bf 23}, 347 (1981).
\bibitem{ref8b}
A. A. Starobinsky, Phys. Lett. B {\bf 91}, 99 (1980).
\bibitem{ref8c} 
A. Linde, Phys. Lett. B {\bf 108}, 389 (1982).
\bibitem{ref8d} 
A. Albrecht and P. Steinhardt, Phys. Rev. Lett. {\bf 48}, 1220 (1982).
\bibitem{ref8e}
A. Linde, ``Particle Physics and Inflationary Cosmology'', Harwood, Chur (1990).
\bibitem{ref8f}
A. R. Liddle and D. H. Lyth, ``Cosmological inflation and large-scale structure'', Cambridge University Press (2000).

\bibitem{ref9a}
Y. Fujii, Phys. Rev. D {\bf 26} (1982) 2580.
\bibitem{ref9b}
L. H. Ford, Phys. Rev. D {\bf 35} (1987) 2339.
\bibitem{ref9c}
C. Wetterich, Nucl. Phys. B {\bf 302} (1988) 668.
\bibitem{ref9d}
B. Ratra and P. J. E. Peebles, Phys. Rev. D {\bf 37} (1988) 3406.
\bibitem{ref9e}
T. Chiba, N. Sugiyama and T. Nakamura, Mon. Not. R. Astron. Soc. {\bf 289} (1997) L5.
\bibitem{ref9f}
P. G. Ferreira and M. Joyce, Phys. Rev. Lett. {\bf 79} (1997) 4740.
\bibitem{ref9g}
P. G. Ferreira and M. Joyce, Phys. Rev. D {\bf 58} (1998) 023503.
\bibitem{ref9h}
E. J. Copeland, A. R. Liddle and D. Wands, Phys. Rev. D {\bf 57} (1998) 4686.
\bibitem{ref9i}
R. R. Caldwell, R. Dave and P. J. Steinhardt, Phys. Rev. Lett. {\bf 80} (1998) 1582.
\bibitem{ref9j}
I. Zlatev, L. M. Wang and P. J. Steinhardt, Phys. Rev. Lett. {\bf 82} (1999) 896.

\bibitem{ref10}
L. Brink, S. S. Kim and P. Ramond, 
JHEP {\bf 0807} (2008) 113 [arXiv:0804.4300 [hep-th]].

\bibitem{ref11}
E. Cotner, A. Kusenko, M. Sasaki and V. Takhistov, JCAP {\bf 10} (2019) 077 [arXiv:1907.10613 [astro-ph.CO]].
\bibitem{ref11a}
Y. B. Zel'dovich and I. D. Novikov, 
Sov. Astron. {\bf 10} (1967) 602. 
\bibitem{ref11b}
S. Hawking, 
Mon. Not. Roy. Astron. Soc. {\bf 152} (1971) 75.
\bibitem{ref11c}
B. J. Carr and S. W. Hawking, 
Mon. Not. Roy. Astron. Soc. {\bf 168} (1974) 399.

\end{thebibliography}
\end{document}